# Internal structure of Pluto and Charon with an iron core

A. AITTA

*Department of Applied Mathematics and Theoretical Physics,*

*University of Cambridge, Wilberforce Road, Cambridge CB3 0WA, UK*

E-mail: A.Aitta@damtp.cam.ac.uk

Pluto has been observed by the New Horizons space probe to have some relatively fresh ice on the old ices covering most of the surface. Pluto was thought to consist of only a rocky core below the ice. Here I show that Pluto can have an iron core, as can also its companion Charon, which has recently been modelled to have one. The presence of an iron core means the giant impact origin calculations should be redone to include iron and thus higher temperatures. An iron core leads to the possibility of a different geology. An originally molten core becomes solid later, with contraction and a release of latent heat. The space vacated allows the upper rock layers to flow downwards at some locations at the surface of the core, and some of the ice above the rock to descend, filling the spaces left by the rock motion downwards. These phenomena can lead to the forces recently deforming the icy surface of Pluto, and in a lesser way, of Charon.

Pluto is a dwarf planet of radius[1,2] $R = 1186$ km, total mass[3] $M = 1.305 \cdot 10^{22}$ kg and mean density 1.868 g/cm$^3$. The basic planetary equations for mass and pressure are needed here. The mass $m(r)$ inside a sphere of radius $r$ is $m(r) = 4\pi \int_0^r \tilde{r}^2 \rho(\tilde{r}) d\tilde{r}$ where $\rho$ is the density. For a body of radius $R$, the total mass is $M = m(R) = 4\pi \int_0^R r^2 \rho(r) dr$. The pressure at radius $r$ and hence depth $h = R - r$ is $P(h) = \int_0^h g(\tilde{h}) \rho(\tilde{h}) d\tilde{h}$ where the gravitational acceleration at depth $\tilde{h}$ is

$$g(\tilde{h}) = \frac{G\, m(R-\tilde{h})}{(R-\tilde{h})^2},$$

with $G$ Newton's constant. When the mass and radius are known, one can find the planet's internal structure easily if it has only two layers: for instance, iron core and rocky mantle. Using integration one can find where their boundary is, assuming one has a good estimate of $\rho(P)$ for both layers. When a planetary body has more layers, one needs to find other constraints or make assumptions about the thickness of the additional layers. Previous works model Pluto and usually also Charon as having a rocky core and icy mantle. Here I demonstrate a multi-layer structure for both bodies: iron core, rocky mantle and icy outermost shell, and how that can generate transport of mass.

Recall how a planet with the size and mass of Pluto looks internally if made only from rock and ice. The rock density as a function of pressure is known only for the Earth[4], and these PREM results are used here even though they may not be precise for Pluto. At the pressures relevant for Pluto, the Earth has three layers of rock: the density is 2.6 g/cm$^3$ between 0.03 and 0.336 GPa, 2.9 g/cm$^3$ between 0.337 and 0.604 GPa, and at higher pressures the density is approximately $3.2872 + 0.032775\, P$, as in Aitta[5]. The ice density as a function of $P$ (in GPa) is estimated as $0.92 + 0.096385\,P$ from fig. 1 in ref. 6. From an initial profile for $P(r)$ one can integrate the density $\rho(r)$ to find the body's mass, and match Pluto's true mass by adjusting the rock-ice boundary radius. Then from the newly obtained density profile one calculates an improved pressure profile $P(r)$ and continues iteratively until $P(r)$ stabilizes. This result is shown in Figure 1. The ice thickness is about 276 km and at the rock-ice boundary the pressure is 0.177 GPa. The pressure at the centre is 1.34 GPa. This is in the range of 1.1-1.4 GPa estimated[7] for a differentiated rock-ice Pluto. The moment of inertia factor is 0.310. But such a layering structure is not expected to generate any interesting dynamics. Thus there is no obvious reason for melting or solidification processes.

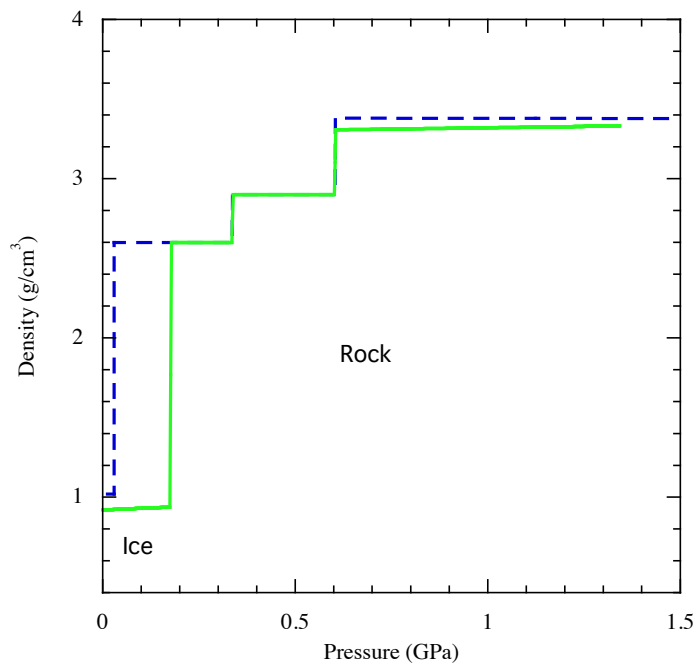

Figure 1. Pluto's internal density profile if made only from rock and ice. The rock density is taken to follow PREM[4] which is the dashed line.

More interesting dynamical events take place if the ice layer is much thicker. The ice is here assumed to reach down to the pressure $P_c = 0.209$ GPa because the melting curve of ice has a minimum there[8]. Then there would be the smallest temperature difference inside the ice layer and the density difference between the surface ice and the ice at this critical point is smaller than the density difference for the other ice phases or liquid. To balance a thick ice layer, Pluto must have in addition to a rocky mantle a small, dense iron core in order to incorporate enough mass. All the terrestrial planets and the Moon are known to have iron-rich cores. Since Pluto is a small body, impurities in its iron core can be neglected. The density of solid iron as a function of pressure, shown at the top of Figure 2, is from Aitta[9]. The core size can again be estimated by iterative integration to get the total mass right. The iteration is continued until the pressure profile stabilizes. The rock-ice boundary is at the radius where the pressure $P = P_c$. The resulting density profile, shown in Fig. 2, has core radius 368 km, the core at pressures between $P = 1.36$ and 2.33 GPa, the ice thickness 320 km, and the moment of inertia factor 0.290.

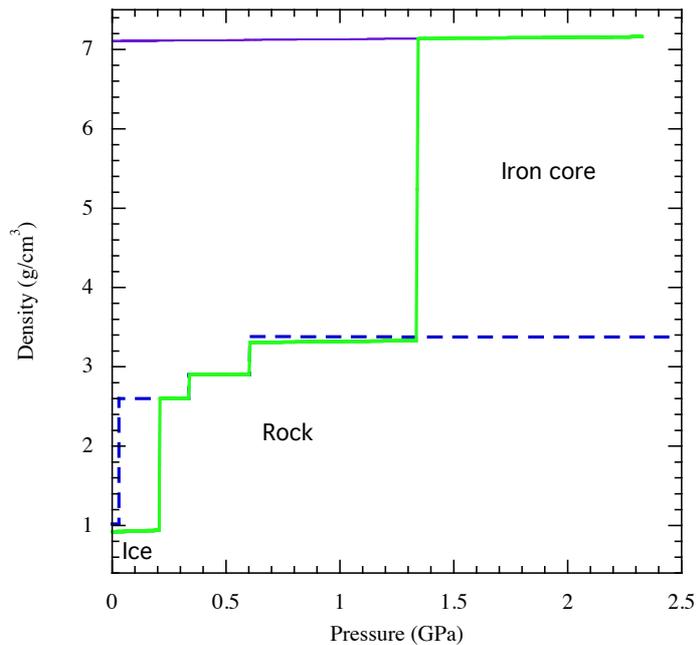

Figure 2. Pluto's density profile consisting of iron, rock and ice. The rock density is taken to follow PREM (dashed line). The iron core density follows the solid iron density curve[9] (top line).

By itself this is a stable internal structure. But very early on Pluto was presumably hotter and the core molten. Molten iron's density is given in ref. 9, too. The radius of a molten core would be 2.2 km larger, but because the core is small, the total radius and mean thickness of ice stay essentially the same.

As a small planet cools, its core solidifies from the core-rock boundary inwards. Typically, the symmetry of a spherical boundary breaks and solidification starts at a particular location (instead of everywhere simultaneously). The core contracts and some of the rocky material descends to occupy the space that becomes available. As a consequence some of the ice at the rock-ice boundary higher up descends too, to fill the space generated by the descending rock. This creates a growing depression on the surface, where the surrounding ice flows down to reduce the height difference.

The total difference in the volume of the molten and solid cores (about 3.8 ·10$^6$ km$^3$) would have caused a major reorganization of material during solidification. It could create a drop of about 7 km in Sputnik Planum (with area around 540 000 km$^2$). The transition would have been gradual, but the edges would have collapsed, generating enormous cracked ice mountains which can be observed on the edges of the plain; the tallest ones are about 3 km high. Gravity would have forced the ice to flow generating a new crater free surface as can be seen in the New Horizons photographs[1].

Charon has radius[1,10] $R = 604$ km, total mass[3] $M = 1.52·10^{21}$ kg and mean density 1.647 g/cm$^3$. Thus it cannot have an ice shell reaching to the same critical pressure. In the absence of this constraint it is here assumed that the core of Charon is comparable in relative size to the core of Pluto, with a scaling by their density ratio 1.134. This gives an estimate for the core radius of 165 km. The iterative integration leads to an ice thickness of 168 km assuming the rocky mantle density is 2.6 g/cm$^3$ as in PREM at the corresponding pressures. The pressure at Charon's centre is only 0.453 GPa, at the iron core-rock boundary 0.258 GPa, and its moment of inertia factor is 0.305. The profile obtained is shown in the Figure 3. Charon has been recently modelled to have an iron core by Rhoden et al.[11] They assume the iron/rock fraction to be 0.29 and then obtain a 200 km radius iron core (with density only 5.5 g/cm$^3$) and a 233 km deep ice-ocean.

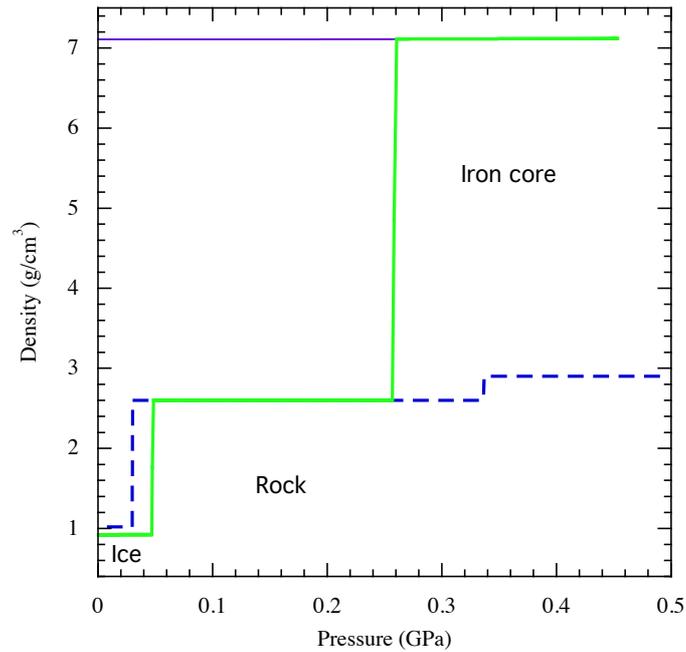

Figure 3. Charon's density profile consisting of iron, rock and ice. The rock density is taken to follow PREM (dashed line). The iron core density follows the solid iron density curve as in Fig. 2.

The core volume change in solidification is for Charon only $3.5 \cdot 10^5$ km$^3$, since the molten core radius is only 1.0 km larger than the solid. Rock and ice descend at some locations also for Charon, but the transport of mass is much reduced. Thus less dramatic surface deformations take place, so dynamically Pluto and Charon are different. A dark depression is visible near Charon's pole as well as a 1600 km long, deep (7 - 9 km in places) and wide (10 to 120 km) canyon. The volume contraction equals this length of depression if it were 8 km deep and 27 km wide everywhere.

A constraint on the depth of Pluto's ice layer is that the iron core needs to be sufficiently substantial to generate a sufficient volume change in the core solidification. The depth considered here is particularly interesting. The melting curve of ice has a minimum temperature $T_c$ = 251K at $P_c$ = 0.209 GPa where the structure changes from hexagonal $I_h$ to ice III. Ordinary ice $I_h$ is less dense than water, but the high pressure phases III, V, VI, VII at the melting curve are denser. The melting curve with its liquid - two solids triple points[8] is shown in Figure 4. The ice on Pluto, which reached originally to $P_c$, will change phase from $I_h$ to ice III when deeper down and under higher pressure. Deeper down it is warmer, too, and the ice melts and expands. This water is less dense than the ice III (or V, VI and VII, but not $I_h$) and is pushed upwards forcing cracks in the ice. Water recrystallizing higher up, at $P_c$ or less, is ordinary ice $I_h$, so another expansion occurs driving more cracks into the ice above.

One can estimate Pluto's temperature profiles when its core was starting to solidify at its surface, and when solidification was just complete. The iron melting temperature[12] at the iron core-rock boundary pressure (for a molten core about to solidify) or at the centre pressure (for a just solidified core) is quadratically connected, as a function of pressure, to the ice critical point ($P_c$, $T_c$) = (0.209 GPa, 251K) and to the surface pressure and temperature (0 GPa, 38 K)[1]. These profiles are shown in Fig. 4. At the present time, the internal temperature is presumably lower. For Charon, the temperature on ice's melting curve for the pressure of its rock-ice boundary is used instead of ($P_c$, $T_c$). These temperatures are so much higher than the existing estimates that one would need to include iron in the giant impact calculations[13].

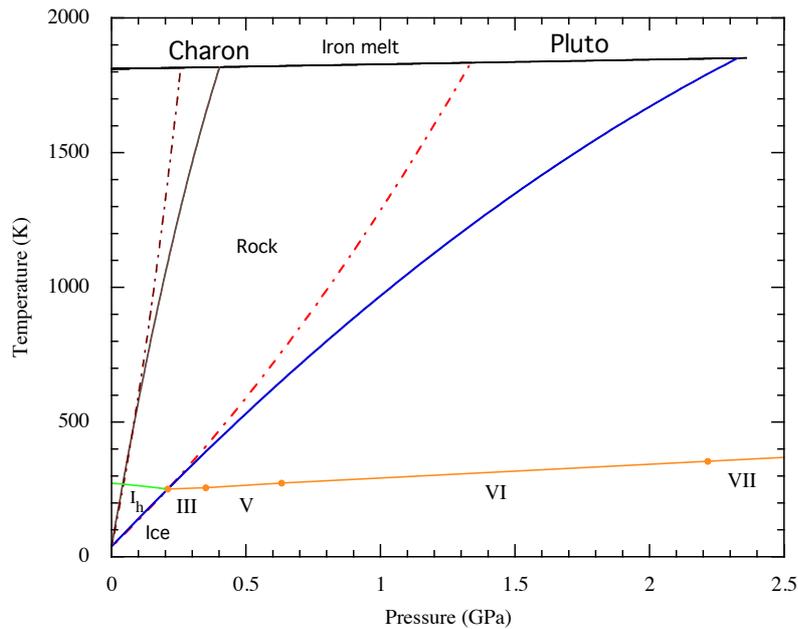

Figure 4. Temperature profiles for the solid parts of Pluto and Charon, with completely molten iron core (dashed dotted) and just fully solidified core (solid curve). The top line is the iron melting curve[12] and the bottom line is the melting curve of ice[8] with Roman numerals for different phases.

In the light of this investigation, and the surface observations by the New Horizons project, it seems inevitable that both Pluto and Charon have iron cores and that their solidification has driven the geological processes seen on the surfaces. Their size difference causes the differences in the amount and nature of the surface deformations.

Acknowledgements: I thank Brenda Frye for encouraging me to work on Pluto's interior now that so much new data is available.